\documentclass[%
 aip,
 jmp,%
 amsmath,amssymb,
reprint,%
]{revtex4-1}

\usepackage{graphicx}
\usepackage{dcolumn}
\usepackage{bm}

\begin{document}

\preprint{AIP/123-QED}

\title[Highly Tunable Hybrid Quantum Dots with Charge Detection]{Highly Tunable Hybrid Quantum Dots with Charge Detection}

\author{C. R{\"{o}}ssler}
 \email{roessler@phys.ethz.ch}
\author{B. K{\"{u}}ng}
\author{S. Dr{\"{o}}scher}
\author{T. Choi}
\author{T. Ihn}
\author{K. Ensslin}
\affiliation{Solid State Physics Laboratory, ETH Zurich, 8093
Zurich, Switzerland}
\author{M. Beck}
\affiliation{Institute for Quantum Electronics, ETH Zurich, 8093
Zurich, Switzerland}
\date{\today}

\begin{abstract}
In order to employ solid state quantum dots as qubits, both a high
degree of control over the confinement potential as well as
sensitive charge detection are essential. We demonstrate that by
combining local anodic oxidation with local Schottky-gates, these
criteria are nicely fulfilled in the resulting hybrid device. To
this end, a quantum dot with adjacent charge detector is defined.
After tuning the quantum dot to contain only a single electron, we
are able to observe the charge detector signal of the quantum dot
state for a wide range of tunnel couplings.
\end{abstract}

\pacs{73.63.Kv, 73.63.Nm, 73.23.Hk }
\keywords{Quantum Dot, Local Anodic Oxidation, Charge Readout, Single Electron}
\maketitle

Quantum dots (QDs) are a playground for quantum engineered devices,
since many system properties like tunnel coupling, energy spacing,
etc. can be controlled and varied. In particular, electrostatically
defined quantum dots, created by local depletion of a
two-dimensional electron gas (2DEG) in an ${\rm Al}_x{\rm
Ga}_{1-x}{\rm As}$ heterostructure, allow to build charge- and
spin-qubits\cite{hay03,pet05}. To this end, both a high degree of
tunability of the confinement potential as well as the capability to
sense the charge state of the QD are needed. By employing
Schottky-split-gates it is possible to tune the local electrostatic
potential in a way that only one electron is left in the
QD\cite{cio00}. Measuring the conductance of a nearby quantum point
contact (QPC) facilitates to determine the charge state of the QD
even if no measurable current flows through the QD\cite{fie93}.
However, electrostatic screening of metal gates between QD and QPC
strongly decreases the readout fidelity as compared to fabrication
techniques without metal gates, like etching or local anodic
oxidation (LAO)\cite{gus09}. But the latter fabrication techniques
have the disadvantage of a low tunability because the confinement
potential is predefined after fabrication. Tackling this issue by
employing a patterned top gate appears to sacrifice the readout
capabilities of LAO defined QDs\cite{sig06}. The combination of
local Schottky-gates with LAO promises to combine highly tunable
confinement potentials with good detector readout fidelity.

\begin{figure}
\includegraphics[scale=1]{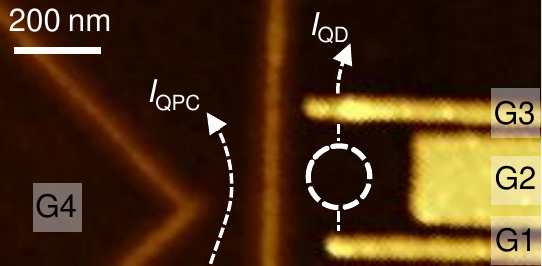}
\caption{\label{fig:sample} (color online) AFM micrograph of the
sample surface (black). The vertical and diagonal oxide lines are
$10\,\rm{nm}$ high and define a QPC in the underlying 2DEG. The 2DEG
area labeled G4 on the left hand side is used to capacitively
control the current through the QPC. Applying voltages to the
$30\,\rm{nm}$ thick Schottky gates G1, G2 and G3 (yellow) defines a
QD.}
\end{figure}

The fabrication is carried out on an ${\rm Al}_x{\rm Ga}_{1-x}{\rm
As}$ heterostructure. The 2DEG resides at the heterointerface,
$z\approx40\,\rm{nm}$ beneath the surface. The 2DEG's sheet density
is $n_{\rm S}=4.9\times10^{15}\,{\rm m^{-2}}$ with a Drude mobility
of $\mu=33\,{\rm m^{2}/Vs}$, as determined in Van-der-Pauw geometry
at a temperature of $T=4.2\,{\rm K}$. After defining the 2DEG mesa
and the outer gate leads via optical lithography (not shown),
$30\,\rm{nm}$ thick Ti/Au gates are deposited via e-beam lithography
(yellow areas in Fig.~\ref{fig:sample}).
An atomic force microscope (AFM) is used to record a topographic
image of the sample's surface. By applying a voltage $V_{\rm
TIP}\sim -30\,\rm{V}$ to the AFM tip at ambient conditions, the
heterostructure is locally oxidized and the underlying 2DEG is
depleted\cite{hel98}. Writing oxide lines (vertical and diagonal
lines in Fig.~\ref{fig:sample}) left of the QD-gates defines a QPC.
In a pioneering work on combining Schottky-gates with
LAO\cite{rog03}, the e-beam lithography was done after LAO. Hence,
the alignment had to be done by e-beam lithography with respect to
pre-defined markers, limiting the accuracy to $\Delta x\sim
50\,\rm{nm}$\cite{rog03}. We find that the accuracy of positioning
the oxide line is limited by the AFM lithography step to $\Delta
x\sim 10\,\rm{nm}$ and can easily be compensated in the experiment
by applying appropriate voltages to the gates. On the right side of
the central oxide line a QD (white dashed circle in
Fig.~\ref{fig:sample}) is defined between gates G1, G2, G3 and the
oxide line.

Applying voltages $V\lesssim -0.6\,\rm{V}$ leads to increased
switching noise which is typical for shallow 2DEG's because of the
small tunnel barriers in growth-direction. This experimental
limitation can be overcome by adjusting the pinch-off voltages of
the gates to values close to $V=0$. Applying positive gate-voltages
during cooldown\cite{pio05} should create a depleting (negative)
potential once the gates are set to zero at low temperature. It
turns out that this so called pre-biased cooldown works very well on
the shallow-2DEG structure employed here. Additionally a bias
$V_{\rm 2DEG}$ can be applied to the QPC-circuit with respect to the
QD-circuit, thereby acting as an in-plane-gate.

\begin{figure}
\includegraphics[scale=1]{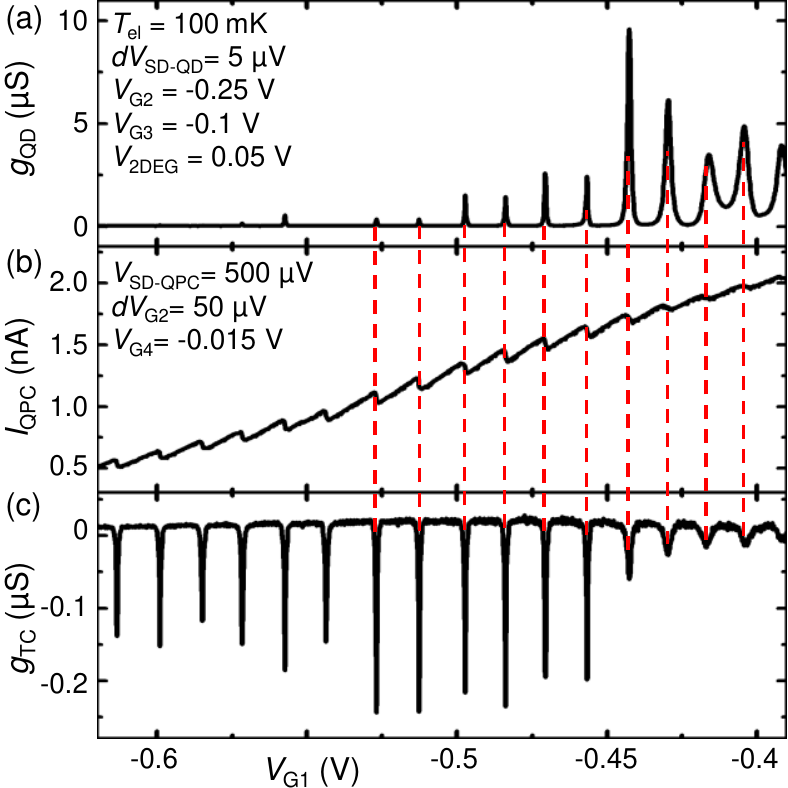}
\caption{\label{fig:readout} (a) Differential conductance of the QD
$g_{\rm QD}$, plotted as a function of gate voltage $V_{\rm G1}$.
Maxima in $g_{\rm QD}$ are Coulomb oscillations. For gate voltages
$V_{\rm G1}<-0.57\,\rm{V}$ (left), no more Coulomb oscillations are
observed. (b) Current through the QPC $I_{\rm QPC}$, plotted for the
same gate voltages. Kinks in $I_{\rm QPC}$ are caused by a change of
the QD occupancy by one electron. (c) Transconductance $g_{\rm
TC}=dI_{\rm QPC}/dV_{\rm G2}$, measured by modulating voltage
$V_{\rm G2}$ and detecting $dI_{\rm QPC}$ with lock-in technique.
Local minima reflect the charge occupancy of the QD.}
\end{figure}

The source-drain current of both circuits is measured at an electron
temperature of $T_{\rm el}\approx 100\,\rm{mK}$ as a function of the
voltages applied to the Schottky gates. Figure~\ref{fig:readout}(a)
shows the differential conductance $g_{\rm QD}$ of the QD-circuit,
measured with lock-in technique as a function of the voltage applied
to gate G1.
Oscillations in $g_{\rm QD}$ indicate that a QD is defined between
the central oxide line and gates G1, G2 and G3, as sketched in
Fig.~\ref{fig:sample}. Since gate G1 also defines one tunnel barrier
of the QD, the amplitude of the Coulomb oscillations decreases
rapidly until no measurable current flows for $V_{\rm
G1}<-0.6\,\rm{V}$. The simultaneously measured current $I_{\rm QPC}$
through the QPC is shown in Fig.~\ref{fig:readout}(b). Stepping
$V_{\rm G1}$ to lower values decreases $I_{\rm QPC}$ due to
capacitive crosstalk between the gate and the QPC. On top of that,
$I_{\rm QPC}$ increases step-like when the occupancy of the QD
changes by one electron. Vertical dashed lines emphasize that the
steps in $I_{\rm QPC}$ coincide with maxima in $g_{\rm QD}$.
Moreover, the QPC is still sensitive to the charge state of the QD
when $g_{\rm QD}$ becomes unmeasurably small. The step height in
Fig.~\ref{fig:readout}(b) is $\Delta I_{\rm QPC}/ I_{\rm
QPC}\approx10\,\%$ at $V_{\rm G1}\approx-0.5\,\rm{V}$. We find step
heights of $5\%$ for strongly coupled QDs and $15\%$ when detecting
charging events of double QDs (not shown). This indicates screening
via nearby electrons in the 2DEG which is reduced when depleting
large areas in order to define a double QD. Table~\ref{tab:readout}
shows typical readout efficiencies of QDs fabricated by different
methods.
\begin{table}
\begin{tabular}{ r || c | c | c | }
Technique: & Schottky& Hybrid & Oxidation \\ \hline
$\Delta I_{\rm QPC}/ I_{\rm QPC}:$ & $1...2\,\%$ & $5...15\,\%$ & $5...40\,\%$ \\
\end{tabular}
\caption{QPC read-out efficiency $\Delta I_{\rm QPC}/ I_{\rm QPC}$
of QDs fabricated by different methods. Values obtained from
\cite{fie93,elz03,gus09,gus06}.}\label{tab:readout}
\end{table}
The step height of the hybrid structures lies in between typical
values of purely Schottky- and purely AFM-defined QDs, with the
spread being due to different sample geometries and QPC pinch-off
slopes.

Figure~\ref{fig:readout}(c) shows the transconductance $g_{\rm
TC}=dI_{\rm QPC}/dV_{\rm G2}$ measured by modulating the voltage
$V_{\rm G2}$ and detecting $dI_{\rm QPC}$ with lock-in technique.
The transconductance minima directly represent the charge occupancy
of the QD\cite{elz03,sch04}. As expected, the peak shape of
transconductance minima and Coulomb oscillations are identical when
the QD is weakly coupled to the leads. In contrast, the peak shape
shows clear deviations in the case of strong coupling. For example,
the rightmost three Coulomb peaks are asymmetric in direct
transport, but symmetric in the transconductance dip. This deviation
has to our knowledge not been observed before in direct transport
and demonstrates the high fidelity of our detector readout.

\begin{figure}
\includegraphics[scale=1]{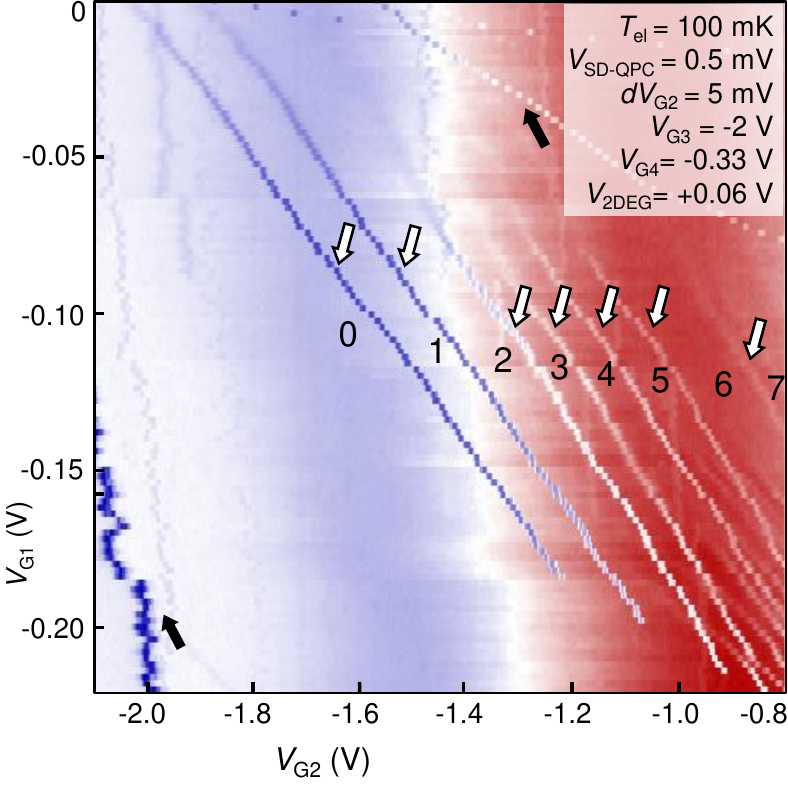}
\caption{\label{fig:transconductance} (color online)
Transconductance $g_{\rm TC}=dI_{\rm QPC}/dV_{\rm G2}$ in false
colors from $g_{\rm TC}=2\times 10^{-9}\,\rm{S}$ (blue) to $g_{\rm
TC}=12\times 10^{-9}\,\rm{S}$ (red), plotted as a function of
$V_{\rm G1}$ and $V_{\rm G2}$. Local minima are caused by Coulomb
resonance of the QD (white arrows) or trapped states in the
environment (black arrows). Between the QD resonances, the electron
number is fixed (labeled from 0 to 7).}
\end{figure}

In order to explore the tunability of our scheme, another sample
with similar geometry is tuned to a very asymmetric configuration
where it is only coupled to the 2DEG underneath gate G1 ($V_{\rm
G1}>-0.3\,\rm{V}$, $V_{\rm G2}<-0.8\,\rm{V}$, $V_{\rm
G3}=-2\,\rm{V}$). Fig.~\ref{fig:transconductance} shows $g_{\rm
TC}$, plotted in false colors as a function of $V_{\rm G1}$ and
$V_{\rm G2}$.
The transconductance is finite and positive throughout the whole
parameter range, indicating that the QPC is neither pinched off nor
insensitive to changes in the local electrostatic potential. Single
resonances with different slopes (marked by black arrows) are caused
by trapped states in the environment, most likely in the doping
layer or in the oxide line. A series of Coulomb resonances (white
arrows) exhibits the same slope of $\Delta V_{\rm G1}/\Delta V_{\rm
G2}\approx4$, indicating that they correspond to charging events of
the same QD. Moreover, the stronger coupling of gate G1 as compared
to G2 confirms that the QD has been "pushed" away from gates G2 and
G3, towards the oxide line and gate G1. Each QD resonance is
characterized by an abrupt end at very negative $V_{\rm G1}$
(bottom) and a washed out regime for less negative $V_{\rm G1}$
(top). This observation is expected from the sample's design, where
gate G1 controls the height of the tunnel barrier between QD and
2DEG. Making $V_{\rm G1}$ more negative and therefore increasing the
tunnel barrier reduces the tunnel rate between QD and 2DEG until it
is comparable to the gate modulation frequency of $f_{\rm
LI}=193\,\rm{Hz}$ and the QD can not compensate the gate modulation
by electron tunnelling. The full width at half maximum
$\rm{FWHM}\approx1\,\rm{meV}$ of the dips in $g_{\rm TC}$ is
dominated by the modulation amplitude $dV_{\rm G2}$. For smaller
tunnel barriers, the leftmost two resonances start to broaden at
gate voltages of $V_{\rm G1}\gtrsim-0.08\,\rm{V}$. This observation
indicates that due to the reduced tunnel barrier height, the tunnel
broadening exceeds $1\,\rm{meV}$, which corresponds to a tunnel rate
of $\Gamma=\rm{FWHM}/h\gtrsim2\times10^{11}\,\rm{Hz}$. Taking these
two values of the tunnel barrier as references, we can estimate the
coefficient relating top gate voltage and tunnel rate to be of the
order of $12\,\rm{mV}$ per decade. This is in good agreement with
tunnel rate measurements on QDs with a comparable Schottky-gate
layout\cite{ota10}.

Observing each QD state over a wide range of tunnel couplings
strongly indicates that after the leftmost QD resonance, the QD is
completely emptied of electrons. This enables us to label the number
of electrons on the QD ("0" to "7" in
Fig.~\ref{fig:transconductance}). The charging energy of the first
two electrons of $E_{\rm C}\sim10\,\rm{meV}$ (determined from the
slope $\Delta E/\Delta V_{\rm G2}=0.5\times \Delta V_{\rm SD}/\Delta
V_{\rm G2}$) is among the largest values reported for laterally
defined QDs. Strikingly, the addition energy of the electronic
states labeled $2$ and $6$ is slightly larger than the adjacent
addition energies. These "magic numbers" are expected from a
two-dimensional confinement potential and have already been
discussed in transport experiments on QDs\cite{tar96,cio00}.
Moreover, the tunnel broadening of the first two electronic states
sets in at $V_{\rm G1}\sim-0.08\,\rm{V}$, whereas the following four
QD states begin to broaden at $V_{\rm G1}\sim-0.11\,\rm{V}$.
Observing the same tunnel broadening for different tunnel barrier
heights indicates different tunnel rates of the involved orbital
states. Again, the observation of tunnel broadened states of a
few-electron QD in the QPC signal demonstrates the capabilities of
the presented device.

In conclusion, we fabricated a QD with adjacent QPC by combining
Schottky-gates with local anodic oxidation. The resulting hybrid
device combines the advantages of both techniques. Reduced screening
of the charge detector facilitates good charge readout and the
employed Schottky-gates demonstrate high tunability of the QD.
Tuning the QD to the few-electron regime, we can detect these charge
states over a range of more than nine orders of tunnel coupling.
Signatures of shell filling effects are observed both in the
excitation energy and in the tunnel rate. Further improvement of the
device geometry and the extension to few-electron double QDs
promises devices with very desirable properties in view of defining
solid state qubits.
%

We acknowledge the support of the ETH FIRST laboratory and financial
support of the Swiss Science Foundation (Schweizerischer
Nationalfonds, NCCR Nanoscience).
%

\end{document}